\documentclass[preprint2,longabstract]{aastex}

\slugcomment{To appear in AJ}

\shorttitle{Observational Study of NGC\,7354 }
\shortauthors{Contreras et al.}

\begin{document}

\title{Observational Study of the Multistructured Planetary Nebula NGC\,7354}

\author{M. E. Contreras, R. V\'azquez}
\affil{Instituto de Astronom\'\i a, Universidad Nacional Aut\'onoma
        de M\'exico \\
        Apdo. Postal 877, 22800 Ensenada, B. C., Mexico}
\email{mcontreras@astrosen.unam.mx, vazquez@astrosen.unam.mx}

\author{L. F. Miranda}
\affil{Instituto de Astrof\'\i sica de Andaluc\'\i a, CSIC \\ 
     C/ Camino Bajo de Hu\'etor 50, E-18008 Granada, Spain}
\affil{Present Address: Departamento de F\'\i sica Aplicada, \\
Facultade de Ciencias, Universidade de Vigo \\ 
      E-36310 Vigo, Spain}
\email{lfm@iaa.es}

\author{L. Olgu\'{\i}n,}
\affil{Depto. de Investigaci\'on en F\'\i sica, Universidad de Sonora\\
Blvd. Rosales Esq. L. D. Colosio, Edif. 3H, 83190 Hermosillo, Son., Mexico}
\email{lorenzo@astro.uson.mx}

\author{S. Zavala}
\affil{Instituto Tecnol\'ogico de Ensenada \\
        Av. Transpeninsular 1675, 22835 Ensenada, B.C., M\'exico}
\email{sazo@astrosen.unam.mx}

\and

\author{S. Ayala}
\affil{Instituto de Estudios Avanzados de Baja California, A. C.\\
       Apdo. Postal 75, 22831 Ensenada, B. C., Mexico}
\email{sayala@ideabc.org}

\begin{abstract}

We present an observational study of the planetary nebula (PN) NGC\,7354
consisting of narrow band H$\alpha$ and [\ion{N}{2}]$\lambda6584$ imaging as well as 
low and high dispersion long-slit spectroscopy and VLA-D radio continuum.
According to our imaging and spectroscopic data, NGC\,7354 has four main structures:
a quite round outer shell and an elliptical inner shell, a collection of low-excitation 
bright knots roughly concentrated on the equatorial region of the nebula and two 
asymmetrical jet-like features, not aligned neither with the shells axes, nor with each other.
We have obtained physical parameters like electron temperature and electron density as well as
ionic and elemental abundances for these different structures. 
Electron temperature and electron density slightly vary throughout the nebula going from 
$\simeq 11,000$ to $\simeq 14,000$\,K, and from $\simeq 1000$ to $\simeq 3000$\,cm$^{-3}$, 
respectively. 
The local extinction coefficient $c_{{\rm H}\beta}$ shows an increasing gradient from South to 
North and a decreasing gradient from East to West consistent with the number of equatorial 
bright knots present in each direction. Abundance values show slight internal variations but
most of them are within the estimated uncertainties. In general, abundance values are in good 
agreement with the ones expected for PNe. Radio continuum data are consistent with optically 
thin thermal emission. Mean physical parameters derived from the radio emission are electron 
density $n_e=710$\,cm$^{-3}$ and $M$(H\,II)$=0.22\,M_\odot$.

We have used the interactive three-dimensional modeling tool {\sc shape} to reproduce the 
observed morphokinematic structures in NGC\,7354 with different geometrical components. Our 
observations and model show evidence that the outer shell is moving faster 
($\simeq$ 35 km\,s$^{-1}$) than the inner one $\simeq$ 30 km\,s$^{-1}$. Our {\sc shape} 
model includes several small spheres placed on the outer shell wall to reproduce to 
equatorial bright knots. Observed and modeled velocity for these spheres lies between the 
inner and outer shells velocity values. The two jet-like features were modeled as two thin 
cylinders moving at a radial velocity of $\simeq$\,60\,km\,s$^{-1}$.
In general, our {\sc shape} model is in very good agreement with our imaging and spectroscopic
observations. Finally, after modeling NGC\,7354 with {\sc shape}, we suggest a possible scenario 
for the formation of the nebula.   

\end{abstract}


\section{Introduction}

Since the recognition that winds from the central stars of planetary nebulae (PNe)
play an important role in the shaping of these objects \citep[e.g.][]{K78,B87}, and that only 
a small fraction of them show circular symmetry, the interest in PN morphology 
has triggered an active field in both theoretical and observational 
astronomy. Many studies have been devoted to classify
\citep[e.g.,][]{B87,SCM92,M96} and to model the basic morphologies observed
(circular, elliptical and bipolar) with noticeable success \citep[e.g.,][]{BPI87,H97}. 
However, high resolution images have shown that the morphologies of PNe are
far from simple. Multiple shells, multipolar structures, highly collimated
outflows, microstructures and peculiar geometries are present in many PNe and 
cannot be explained with simplistic models \citep[e.g.,][]{ST98,MAVG06,MPG09}. 
Nowadays, it is accepted that highly collimated outflows play a crucial role
in the shaping of PNe \citep{ST98}. Nevertheless, other physical processes 
are probably present, so that most likely the shaping of PNe is a result of many
processes acting at the same time \citep[e.g.,][]{BF02}. In order to
understand complex PNe, the first step is to identify the structural
components present and to define their nature. High-resolution, spatially resolved
spectroscopy combined with narrow-band imaging has demonstrated to be a
powerful tool to disentagle the structural components in PNe, to infer their
nature and to constraint the ejection processes involved in their 
formation \citep{MS92,V99,V08,G08,V98,L00}.

Although at first glance NGC\,7354 looks like an elliptical PN, extense H$\alpha$, 
[\ion{O}{3}]$\lambda5007$, [\ion{N}{2}]$\lambda6584$ and mid-infrared imaging and
spectroscopic data have revealed that it possess a more complex structure
\citep{S83,B87,H97,Ph09}. In particular, \citet{B87} described this nebula as
consisting of an inner halo, a thin bright rim, two spike-like tails, and
low-excitation patches projected onto the rim-halo interface. The large-scale
structures observed in NGC\,7354 are qualitatively well described using the
interacting winds theory \citep[e.g.][]{BPI87,M95,H07} but no deep analysis
has been carried out for the small-scale structures and their relationship with the
large-scale ones. 
 
In this work we present a detailed observational study of the morphology,
internal kinematics, and physical and chemical properties of NGC\,7354. These 
data allow us to discuss and model each of the components present in the
object and to suggest a possible scenario for the formation of this nebula.

\section{Observations and Results}

\subsection{Optical Imaging}

Narrow-band images of NGC\,7354 were obtained on 1997 July 24 with the Nordic Optical
Telescope (NOT)\footnote{The NOT is operated on the island of La Palma jointly by Denmark, 
Finland, Iceland, Norway, and Sweden, in the Spanish Observatorio del Roque de los
Muchachos of the Instituto de Astrofisica de Canarias.} and the HiRAC camera equiped with 
a Loral CDD of 2048$\times$2048 pixels and a plate scale of $0\rlap{.}{''}11$ pixel$^{-1}$.
Two narrow-band filters were used: [\ion{N}{2}]$\lambda6584$ ($\Delta \lambda=10$\,\AA) and 
H$\alpha$ ($\Delta \lambda=10$\,\AA). The exposure time for both filters 
was 900s. Seeing was about $0\rlap{.}{''}9$ during the observations.
  
Figure~\ref{fig1} shows a mosaic of our H$\alpha$ and [\ion{N}{2}] images, 
including unsharp masking images in both filters constructed to show up both 
the large- and small-scale structures in the nebula. In these images, we can
identify the main structures previously described by other authors: the outer
shell, the elliptical inner shell, the bright equatorial knots, and the two jet-like 
features. In the following we will describe in more detail each of these structures as 
well as new morphological details that have not been previously mentioned. 

The outer shell looks like a round envelope in the high-contrast images but it
appears as a faint cylindrical structure in the low-contrast images. Its size is
$\simeq 33''\times29''$ with the major axis oriented at
position angle PA\,$\simeq15^\circ$. The outer shell is brighter in
H$\alpha$ than in [\ion{N}{2}] although in [\ion{N}{2}] several bright knots
are observed at the edges of it. 

The inner shell present an elliptical shape with its major axis oriented at PA
$\simeq30^\circ$ and a major and minor axis length of $\simeq$ 21$'$ $\times$ 16$''$, 
respectively. This structure is noticeable fainter in [\ion{N}{2}] than in
H$\alpha$. Moreover, the regions along the minor axis are particularly bright
in H$\alpha$. A closer inspection of the images shows that the polar regions
deviate from the elliptical shape and appear as two bubbles. This is particularly 
noticeable in H$\alpha$. We will refer to these regions of the inner shell 
as the polar caps . 

The bright equatorial knots are observed in [\ion{N}{2}] but not in
H$\alpha$. They are mainly concentrated in two groups, East and West, and
along the equatorial plane of the outer and inner shells. 

The two jet-like features are bright in [\ion{N}{2}] but much fainter in
H$\alpha$. The northern feature is oriented at PA$\simeq13^\circ$ and extends
$\simeq7''$. The southern feature is oriented at PA $\simeq205^\circ$, extends 
$\simeq11''$, and appears narrower than the
northern one. We note that the orientation of these two features does not
coincide with each other nor with the orientation of the outer and inner shells.

In order to improve the view of NGC\,7354, we retrieved a Hubble Space Telescope (HST)
image from the MAST Archive\footnote{Some of the data presented in this paper were 
obtained from the Multimission Archive at the Space Telescope
Science Institute (MAST). STScI is operated by the Association of Universities for Research 
in Astronomy, Inc,, under NASA contract NAS5-26555. Support for MAST for non-HST data provided
by the NASA Office of Space Science via grant NAG5-7584 and by other grants
and contracts.} (Proposal ID:~7501; P.I.: A. Hajian; date of observation:~1998
July 21; filter F658N; exposure time 1000\,sec). Fig.~\ref{fig2} shows
this image. The morphology of the outer and inner shells is similar to that
observed in ground based images (Fig.\,1). The jet-like freatures appear as
cometary tails with a bright knot facing the central star and faint tails
directed outwards. They present a knotty structure, particularly the southern
one. The HST image resolve the bright equatorial knots into a series of small 
knots and filaments embedded in diffuse emission.

\subsection{Radio continuum}

The $\lambda3.6$\,cm radio continuum observations were made with the Very Large Array (VLA)
of the NRAO\footnote{The National Radio Astronomy Observatory (NRAO) is operated by 
Associated Universities Inc., under cooperative agreement with the National
Science Foundation.} on 1996 May 31 in the DnC configuration. The standard VLA continuum mode 
with a bandwidth of 100\,MHz and two circular polarizations was employed. Phase center was
set at RA(2000.0)=22$^{\rm h}$40$^{\rm m}$20\fs1, DEC(2000.0)=+61\arcdeg17\arcmin06\farcs0. 
Flux calibrator was 3C48 (adopted flux density 3.3 Jy) and phase calibrator
was 0019+734 (observed flux density 1.0 Jy). Total on-source integration time was 30 minutes.
The data were calibrated and processed using standard procedures of the 
Astronomical Image Processing System (AIPS) package of the NRAO.

Fig.~\ref{fig3} shows an uniform-weighted map of NGC\,7354. The
emission presents a circular morphology and extends $\simeq$42\arcsec\,in
diameter. Two emission maxima are observed separated by $\simeq$10\arcsec\,and 
oriented at PA$\simeq100\arcdeg$. These emission maxima coincide with the 
H$\alpha$ bright regions observed along the minor axis of the inner shell. 

From our data we derive a peak flux density of 72\,mJy\,beam$^{-1}$  at 
position RA(2000.0)=22$^{\rm h}$40$^{\rm m}$20\fs44,
DEC(2000.0)=+61\arcdeg17\arcmin06\farcs8, and a total flux density
of $502\pm4$\,mJy. We note that our map is similar to that obtained 
by \citet{T74}, being the flux density values obtained in both works consistent with 
each other and with optically thin thermal emission. 
Adopting a distance of 1.5\,kpc for the nebula \citep[][see Sec. 3.1]{S86} and following the 
formulation by \citet{MH67}, we derive a mean electron density of $\simeq$ 710\,cm$^{-3}$ and an 
ionized mass of $\simeq$ 0.22\,$M_\odot$.

\subsection{Low Resolution Spectroscopy}

Low resolution, long-slit spectra were obtained with the Boller \& Chivens spectrograph
mounted on the 2.1\,m telescope at the San Pedro M\'artir Observatory 
(OAN-SPM)\footnote{The Observatorio Astron\'omico Nacional at San
Pedro M\'artir (OAN-SPM) is operated by the Instituto de Astronom\'{\i}a of the Universidad 
Nacional Aut\'onoma de M\'exico.} during three observing runs: 2002 June 14, 2002 August 7, 
and 2002 December 10 and 11. A CCD SITe with $1024\times1024$ pixels was used as a detector. 
We have used a 400 lines/mm dispersion grating and a slit width of 2\arcsec\,giving a
spectral resolution (FWHM) of 7\AA.  Spectra reduction was performed using
IRAF\footnote{The Image Reduction and Analysis 
Facility (IRAF) is distributed by the National Optical Astronomy Observatories, which are 
operated by the Association of Universities for Research in Astronomy, Inc., under 
cooperative agreement with the National Science Foundation.} 
standard procedures. We have used four slit positions to cover specific regions of the
nebula. Fig.~\ref{fig4} shows these slits, labelled A to D, and the regions extracted from 
each long-slit spectrum overplotted on the unsharp masking [\ion{N}{2}] image. 
Regions are denoted by a letter refering to the slit position followed by a sequence number 
along the slit. In total, 21 regions in NGC\,7354 have been analyzed. 

Table\,1 presents dereddened line fluxes, observed H$\beta$ flux and the  
logarithmic extinction coeficiente $c{_{\rm H\beta}}$ for each region, the last has been 
derived from the observed H$\alpha$/H$\beta$ ratio assuming case B recombination.  
The fluxes have been dereddened aplying the extinction law by \citet{CCM89}. Electron 
temperature ($T_e$) has been derived from the [\ion{O}{3}] and/or [\ion{N}{2}]
emission lines, while electron density  ($N_e$) has been derived from the [\ion{S}{2}], 
[\ion{Cl}{3}] and/or [\ion{Ar}{4}] emission lines, in both cases using the Five-Level Atom 
Diagnostic Package NEBULAR \citep{dRDH87,SD94} from IRAF. Derived physical parameters and
their estimated errors are shown in Table~\ref{tbl-2}.  Error
estimates take into account the readout and photon noise and they are
propagated along the calculation of physical quantities.  Whenever possible,
$N_e$ values were calculated using the value of $T_e$ derived from 
the corresponding high- or low-excitation ion, [\ion{O}{3}] or [\ion{N}{2}], respectively.  

Extinction within the nebula, as indicated by $c_{\rm H\beta}$, shows a
slight increase from South to North with values ranging from $\simeq$\,1.7 to
$\simeq$\,2.4, respectively. Relatively high values of $c_{\rm H\beta}$ $\simeq$
2.4 are found in the low-excitation equatorial knots. These values can be
compared to those in the surroundings of the knots where the extinction
decreases up to $c_{H\beta}$ $\simeq$ 1.7. This result suggests that the
equatorial knots are denser and/or contain more dust than the rest of the nebula.

Electron density slightly increases inwards from the outer shell (regions B1 and C6) with 
values of $\simeq$~1\,000\,cm$^{-3}$ to the inner regions of the inner shell (region
B4) where values of $\simeq$ 2\,400\,cm$^{-3}$ are found. The bright
equatorial knots are clearly denser that both the outer and inner shells
with an average density of $\simeq$\,2\,600\,cm$^{-3}$. For the jet-like
features, the electron density is relatively low with values around 
$\simeq$\,1\,300\,cm$^{-3}$.

Electron temperature also seems to slightly increase from $\simeq$ 13,000\,K at the walls 
of the outer shell (regions B1, B6 and B7) to $\simeq$ 15,000\,K at the center of
the inner shell (region B3). In the bright equatorial knots, electron
temperature ranges from $\simeq$~10,000 to $\simeq$\,12,400\,K. For the northern
jet-like feature, no [\ion{O}{3}] and [\ion{N}{2}] lines 
were detected with a good signal-to-noise; in the southern jet-like feature,
we have obtained an electron temperature of $\simeq$ 12,000\,K (regions A5 and
D3). 

Ionic and elemental abundances values are listed in Tables 3 and 4,
respectively. They have been obtained with the task IONIC in IRAF and using
the ionization correction factors by \citet{KB94}, respectively. 
Table~\ref{tbl-4} also lists abundances in other objects for comparison purposes.
Small abundance variations, within the estimated uncertainties, are observed throughout 
NGC\,7354. In general, we found that all our abundance determinations are consistent with 
those of PNe, see Table~\ref{tbl-4}.  Since neither He nor N overabundance is observed, 
as in the case of Type\,I PNe, we can say that it behaves like a Type\,II PN.
 
Elemental abundances for NGC\,7354 have been reported by several authors \citep{H97,MV02,P04,
S06}. However, a comparison of our abundance estimations with those from the literature is not
straightforward since we have obtained abundance values for several slit positions and specific 
regions along them. In the case of \citet{H97}, they report abundance values of six regions 
along a slit position very similar to our slit A. Their abundances tend to be two or three times
higher than ours in the case of O/H, N/H and S/H but lower in He/H and similar in Ar/H. 
Comparison with other studies can only be done through average values along each of our slit 
positions or the total average obtained from all our studied regions. We have compared our total 
averaged O/H, N/H, S/H, Ar/H abundances with those from \citet{P04} and \citet{S06}. In both 
studies, their abundances seem to be lower than ours but in the case of \citet{S06} this difference 
may be due to tha fact that they consider the PN as a whole and excluded special features in the 
nebula.

\subsection{High Resolution Spectroscopy}

High-resolution, long-slit spectra were obtained in 2002 July 15 to 17 and 2007
July 10 to 17 with the Manchester Echelle Spectrometer (MES; \citep{M03}) mounted on the 
2.1\,m telescope at San Pedro M\'artir Observatory (OAN-SPM). A Site CCD with 
1024$\times$1024 pixels was used as a detector. Binnings of $1 \times
1$  and  $2 \times 2$ were used in 2002 and 2007, respectively. Slit width was
1.6\arcsec\, and the achieved spectral resolution (FWHM) is 12 km
s$^{-1}$. The slit was oriented North-South and centered at several right
ascensions across the nebula, except in those slits that covered the jet-like
features, which have been centered on the central star and oriented at PAs
13$^{\circ}$ and 25$^{\circ}$. Figure\,5 shows the used slit positions,
numbered from 1 to 8, superimposed on the unsharp masking  [\ion{N}{2}]
image. Data reduction was carried out with the IRAF package.

Grey-scale, position-velocity (PV) maps, derived from the eight long-slit
spectra, are shown in Fig.~\ref{fig6} for the H$\alpha$, [\ion{N}{2}]$\lambda6584$, and
\ion{He}{2}$\lambda6560$\ emission lines. From the long-slit spectra, we
derived a heliocentric systemic velocity of $-42\pm2$\,km\,s$^{-1}$ for
NGC\,7354, in excellent agreement with $-41\pm2$\,km\,s$^{-1}$ deduced by \citet{S83}. 
Through this paper, we will consider the systemic velocity as the origin for quoting internal
radial velocities and the declination of the central star as the origin for quoting distance
measurements. The PV maps allow us to recognize the structural components that
have been identified in the direct images. In the following, we will describe
the spatio-kinematical properties of these components.

The [N\,{\sc ii}] emission from the outer shell is recognizable at slit positions 
1,2,3,5, and 6. The emission is very faint and in the central nebular regions it appears
superposed by the stronger emission from the inner shell. Maximum radial 
velocity of $\simeq$ 35 km\,s$^{-1}$ is observed at the center and decreases with 
distance to the central star. Emission from the outer
shell can be recognized in the H$\alpha$ line (Fig.\,6) by its spatial
extend. However, the large thermal width and, probably, low expansion velocity
in this line do not allow us to obtain detailed kinematic information. The
outer shell cannot be recognized in the long-slit spectra of the
\ion{He}{2}$\lambda6560$ line.

The inner shell can be identified at all slit positions in the three
lines. The emission feature of the inner shell appears as a velocity ellipse
in the PV maps. The size of the velocity ellipse is maximum at slit 8 with
values of 30$''$ in H$\alpha$ and  [\ion{N}{2}], and 20$''$ in
\ion{He}{2}. Maximum velocity splitting is observed at the center of the
nebula (slits 3, 7, and 8) and amounts 60, 56 and 48 km\,s$^{-1}$ in
H$\alpha$, [\ion{N}{2}], and \ion{He}{2}, respectively. This implies expansion
velocities for the inner shell of 30, 28 and 24 km\,s$^{-1}$,
respectively. Since we estimate an error of $\leq 1$\,km\,s$^{-1}$
in our velocity measurements, these results show that the \ion{He}{2} emission is 
confined to a slow expanding, inner thin layer of the inner shell while the 
[\ion{N}{2}] traces the outer layer that expands faster.   

It is worth noticing the very faint and wide emission observed in [\ion{N}{2}]
(slit positions 2, 3, 7 and 8) located at about 12\arcsec\,and 8\arcsec\,to 
the North and South of the central star, respectively. The velocity of this faint emission 
spreads from $\simeq -45$ to $\simeq33$\,km\,s$^{-1}$. Comparing our [\ion{N}{2}] 
direct image and these particular PV maps, we identify these weak emissions with the 
two polar caps located on the main symmetry axis of the inner elliptical shell.

\begin{figure*}
\epsscale{1.6}
\plotone{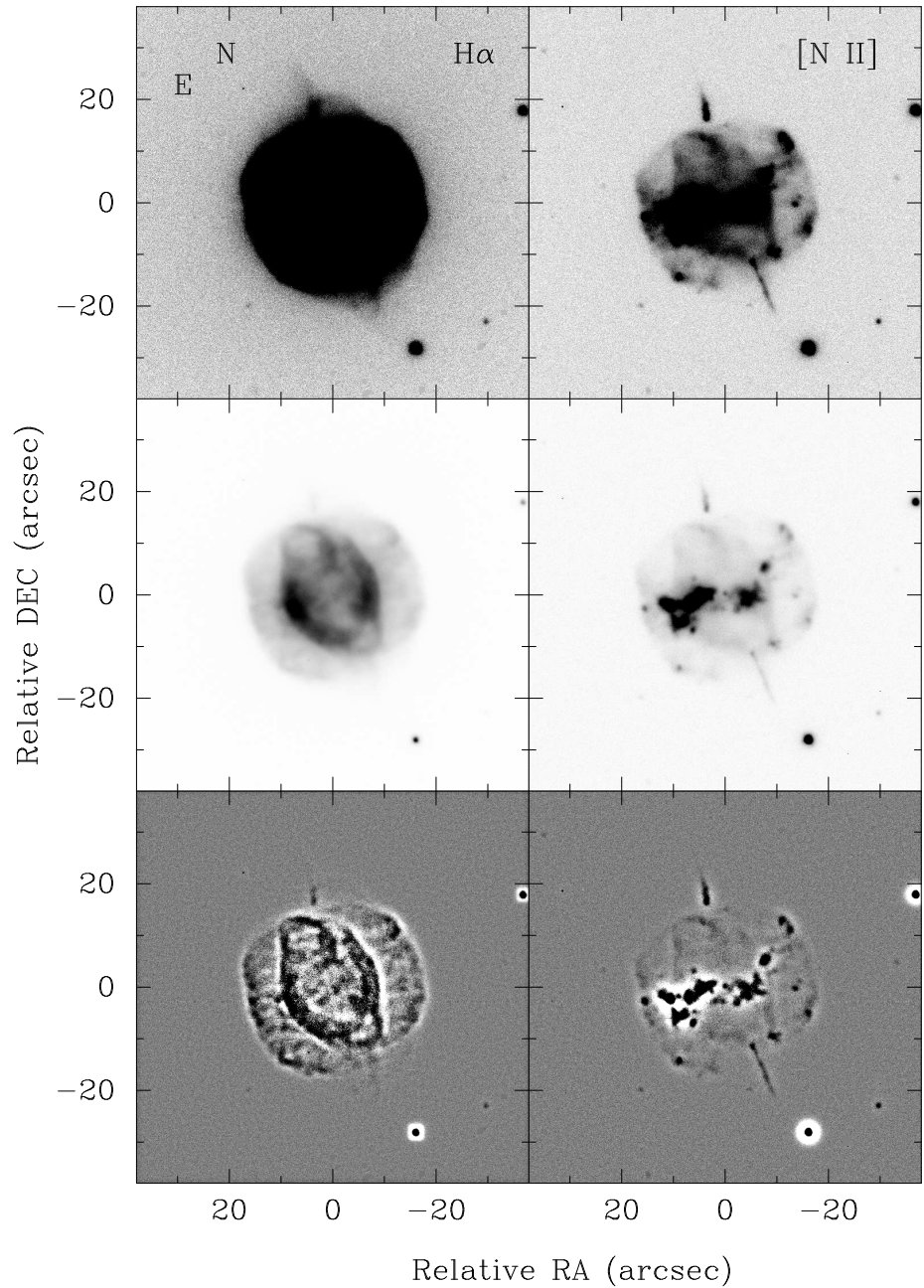}
\caption{Grey-scale representations of the images of NGC\,7354 in the light 
of H$\alpha$ (left panels) and [\ion{N}{2}]$\lambda6584$ (right panels). Two
different contrasts have been used to show up the faint and bright nebular
structures.  Unsharped masking images are also present (lower row) in order to
see details of the nebula.\label{fig1}}
\end{figure*}

\begin{figure*}
\epsscale{1.5}
\plotone{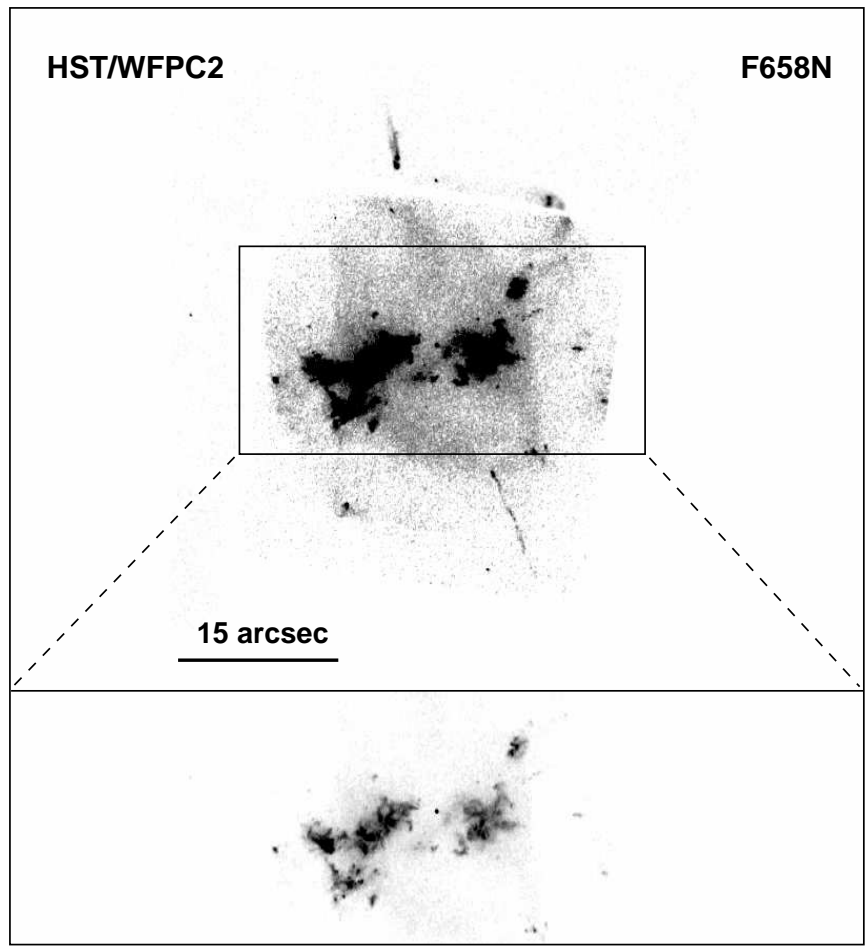}
\caption{Grey-scale representation of the F685N HST/WFPC2 image of NGC\,7354. 
The same features described from ground-based images can be identified here. 
The lower frame shows a lower contrast view of the equatorial region of 
NGC\,7354.\label{fig2}}
\end{figure*}

\begin{figure*}
\epsscale{1.7}
\plotone{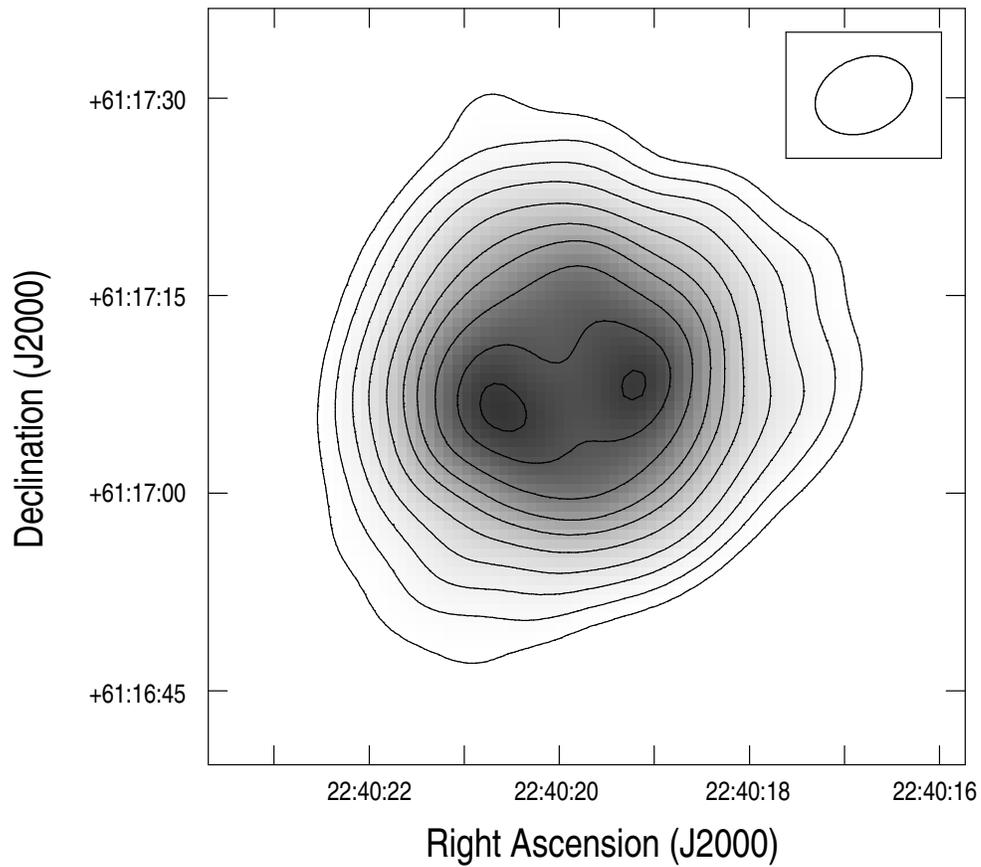}
\caption{Grey-scale and contour uniform-weighted map of NGC~7354 at 3.6\,cm radio 
continuum. Contour levels are 8, 18, 35, 60, 100, 150, 200, 270, 350, and 400 times 
144~$\mu$Jy~beam$^{-1}$, the rms noise in the map. The half-power beam
width ($8\farcs0\times6\farcs5$, PA~$-86\arcdeg$) is shown in the upper right 
corner.\label{fig3}}
\end{figure*}

\begin{figure*}
\epsscale{1.6}
\plotone{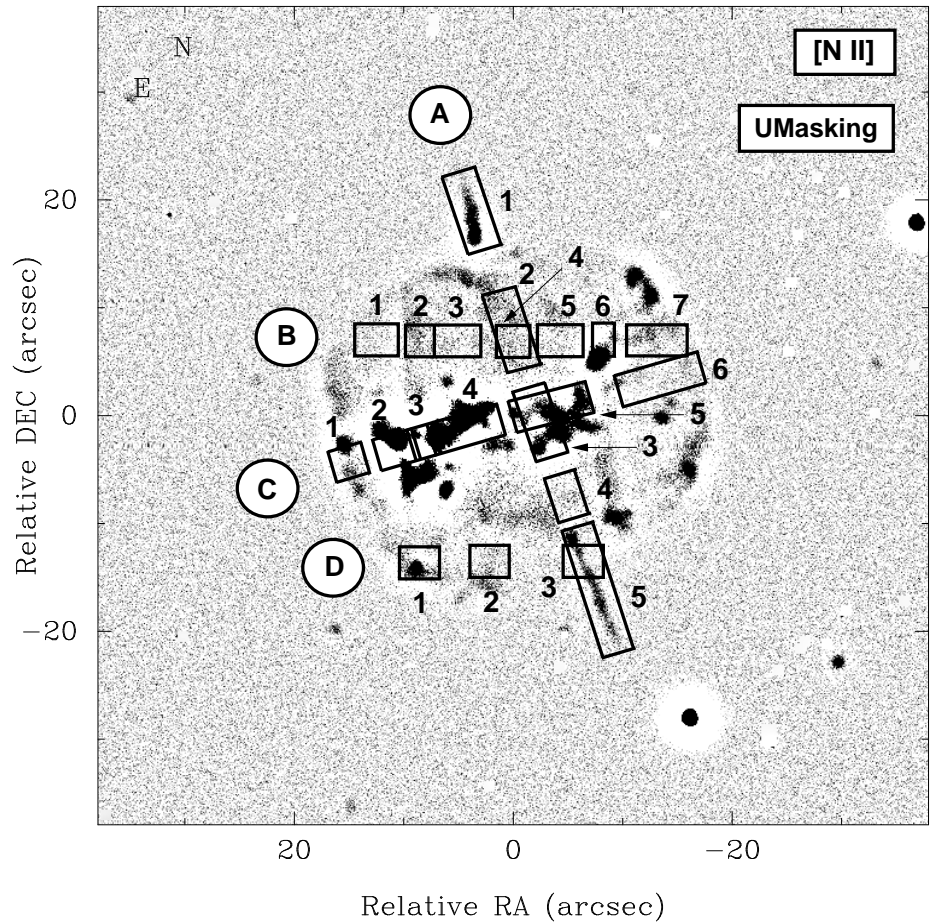}
\caption{Slit positions (A,B,C,D) used for low resolution spectroscopy superimposed
on the unsharped masking [\ion{N}{2}]$\lambda6584$ image. Along each slit position 
several regions (1,2,3,...)  were selected to analyze specific structures of
the nebula.\label{fig4}}
\end{figure*}

All slit positions in our [\ion{N}{2}] PV maps, show the presence of at least one of the 
bright low-ionization knots identified in our optical [\ion{N}{2}] images (Fig.~\ref{fig1}, 
right panel). In all these PV maps we can see that most of them are located very close to the
zero position line, i.e. almost on the equatorial plane of the nebula. Although these PV maps 
show that emission arising from different bright knots may be mixed due to projection 
effects, we can estimate that their expansion velocity ranges between the inner and outer 
shell expansion velocities, 28\,km\,s$^{-1}$ to $\simeq$\,40\,km\,s$^{-1}$, with only two of 
them having lower velocities, $\simeq$\,24\,km\,s$^{-1}$. A final remark on the low-ionization 
knots is that in our [\ion{N}{2}] PV maps, slit positions 1 and 6, we can see two weak spots 
of emission almost simetrically located at about 14\arcsec\,to the North and South from the 
central star, respectively. Both emission spots show negative and very low velocities of 
$\simeq\,-5$\,km\,s$^{-1}$. We identify these two spots of emission as coming from two bright 
knots that, in projection, appear to be located on the outer shell wall (see Fig.~\ref{fig1}, 
[\ion{N}{2}] images).

Although [\ion{N}{2}] emission from the two jet-like features can be identified at slit
positions 2 and 5 (Fig.~\ref{fig6}), it can be better analyzed at slit positions 7 and 8
which were specially selected to cover these structures. In the corresponding PV maps we 
can see the conspicuous emission arising from the two jet-like features. While the North 
jet-like feature emission is slightly redshifted, the South feature is slightly blueshifted. 
Both features show a small radial velocity, between 0 and 5 km s$^{-1}$ which suggest that 
they are moving almost in the plane of the sky.

\section{Discussion}

\subsection{Morphokinematic structure and modeling}

As we have described, NGC\,7354 shows four different structures: two large scale 
structures, the outer and the inner shells, the last having two polar caps located on its 
symmetry axis; a number of low-ionization bright knots lying about the equatorial plane, and 
two jet-like features located close to the North and South of the nebula, slightly inclined 
respect to the inner shell symmetry axis. Our high dispersion spectra 
indicate that (a) the inner shell is expanding with a velocity which is smaller than that of 
the outer shell, (b) most of the bright equatorial knots are moving at velocities closer to 
the outer shell, and (c) the jets' projected velocity is very small. 

In order to test the overall morphology and kinematic structure of the nebula we have used 
the interactive three-dimensional (3D) modeling tool {\sc shape} Ver. 2.0 \citep{SL06}. 
{\sc shape} produces synthetic images and PV diagrams which can be compared directly with our 
observed CCD images and PV maps. We have considered two ellipsoidal structures (inner and 
outer shells), two spherical sections (polar caps), several small spheres with different
diameter (bright knots) and two geometrically thin cylinders of 2$''$ in diameter (jet-like 
features). All these geometrical structures were slightly modified (smoothered, lengthened and/or 
rounded) in order to match the overall looking of our [\ion{N}{2}] images. Besides, all the 
structures were placed on the proper position to get the corresponding observed radial 
velocity from our PV maps. Fig.~\ref{fig7} shows all the geometrical components used to
construct our {\sc shape} model.

 Each one of the main structures was analized separately in order to obtain its spatial 
velocity independently. Due to projection effects we are unable to distinguish neither if the 
low-ionization knots are located between the inner and outer shell nor if they are lying on 
the foreground or on the background side of the nebula. Thus, we have assumed that 
the bright knots are located on the outer shell wall, according to their individual velocity 
shown in our PV map. It is important to remark that the inclination angle of the 
different structures is unknown and {\sc shape} does not solve it. Final geometrical and
kinematic parameters are shown in Table~\ref{tbl-5}. Some of the inner shell basic parameters 
can be compared to those found by \citet{H07}  with an extended prolate ellipsoidal shell model.
While our derived inclination and position angles are consistent with the ones determined by 
\citet{H07}, our expansion velocity  value is larger than theirs. However, as they have 
found in all the cases examined in their study, the [\ion{N}{2}] gas has a larger expansion 
velocity than the [\ion{O}{3}] gas. Then, since we have determined the expansion velocity
through the [\ion{N}{2}] line emission, a difference in expansion velocities is expected.
Distance estimations found in the literature lie around 1.2\,kpc \citep{CKS92,Z95,Ph04} being
1.5\,kpc the largest estimation \citep{S86} and 0.88\,kpc the smallest one \citep{D82}. 
We have adopted the largest value of 1.5\,kpc in order to derive an upper limit for the 
kinematical ages. However, using a distance value of 1.2\,kpc implies a decrease in 
the kinematical age estimation of only 20\%. But most important, distance uncertainty does
not modify the proposed chronological sequence of structure formation in our model.

From this model, we have obtained synthetic images and PV diagrams
which strongly resemble the observed ones, see Fig.~\ref{fig8} for an example of this good 
match. Thus, from the proposed 3D model we have not only reproduced qualitatively the overall 
morphology of NGC\,7354 but we have also quantitatively reproduced the kinematic behaviour of 
each structure.

\subsection{Formation of NGC\,7354}

At present, binary star models have been succesful in explaining the origin and process 
responsible for the equatorial density enhacement required in the interacting stellar wind 
(ISW) model, or rather the generalized ISW (GISW) model, to explain early- and middle-type 
elliptical and even bipolar large scale structures observed in PNe 
\citep[e.g.][]{BPI87,M95,MF95}. On one hand, common envelope (CE) evolution in close binaries
has been proposed to be responsible for the expanding slow wind torus needed in 
GISW \citep[e.g][]{RL96,F96,TT96,S98}. On the other hand, accretion disks in binary systems 
seem to account for the narrow waist bipolar morphologies and even the jets present in PNe. 
Details may vary on each binary model in order to explain individual morphologies and 
structures, but in all of them the main idea is that of a jet launched by the central star, 
or its companion \citep[][among others]{SR00,GAF04,S07,D08,AS08}. 

 Since at first glance NGC\,7354 looks like a rather elliptical nebula, one
may think that the single star models (ISW or GISW) can explain the presence of the outer
and inner shells, and even the formation of the two jet-like features
observed in our [\ion{N}{2}] image \citep{BPI87}.  However, it is unable to explain why all 
the different structures show different position angles (PA's) on the sky. This 
characteristic may indicate that the direction of ejection varies with time, and more likely 
that all the features were formed as independent events. Although nowadays there is no 
evidence of NGC\,7354 possessing a binary nucleus, one may turn around and take a look at 
binary star models to try to understand in a consistent way the observed morphology of 
NGC\,7354. 

 It has been proposed that the GISW model coupled with the predictions of the CE evolution 
theory can account for the formation of elliptical and even bipolar PNe morphologies 
\citep{SL94}. Then, we could succesfully explain the outer and inner shells in NGC\,7354 
based on this mixed model in the following way. After the CE phase has ended, i.e. the 
envelope has been ejected, the binary may become closer, mass transfer take place, an 
accretion disk is formed and a jet may be launched \citep{SL94}. If we assume that the 
subsequent evolution is similar to those described in the models for pPNe \citep{SR00,LS03,
D08,AS08} it is possible that a jet arise from an accretion disk. Then, the accretion disk 
theory would be able to explain also the two jet-like features observed in NGC\,7354. 
Moreover, some of these models even propose the formation of an expanding dense ring in the 
equatorial plane \citep{SR00} which in the case of NGC\,7354 might be related to the 
equatorial bright knots. As the nebula evolve, a combination of the processes and mechanisms 
just mentioned might take place.

 With this in mind, we suggest a possible qualitative scenario for the formation of the main 
structures present in NGC\,7354. A slow wind is lost by a central binary system ongoing a CE
phase, producing the mild elliptical outer shell. When the AGB fast wind begins, it
finds an equatorial density enhaced enviroment forming the bright elliptical rim or inner
shell. At this point the CE has been completely ejected and mass transfer from the secondary 
(main-sequence star) into the primary (white dwarf) will form an accretion disk and 
eventually a couple of jets will be launched. Once the jets have appeared, they will ``push''
the polar ends of the inner shell, detaching them from it. Eventually, the two jets would 
break the caps and escape from the main body of the nebula. Along with the processes 
mentioned, the binary system may be precessing and each of the structures would show a 
different PA on the sky. In the case of the two jets, precession may cause that they impinge 
on the polar caps with a certain angle, i.e. not along the symmetry axis of the caps, 
breaking them apparently on the ``base'' of them. According to our derived kinematical ages, 
the chronological sequence of formation seem to be consistent with the above description. 
The outer shell is the oldest structure present in the nebula with an age of about 2500\,yr. 
The next younger structure would be the inner elliptical shell with $\simeq$1600\,yr. 
Kinematical ages for the two jet-like features indicate that they are coeval and both were 
formed almost at the same time as the inner elliptical shell. However, we should notice that 
the age was derived assuming that they have been moving with the same velocity (60 km\,s$^{-1}$) 
since their launch. According with the theory of formation of jets this low initial velocity would 
be very unlikely, since they arise from accelerated flowing material. Then, more likely the 
jets possessed a higher initial velocity but they have been restrain along its way. Maybe 
the main cause of this decceleration was the interaction of the jets with the polar caps.

Regarding the equatorial low-excitation bright knots, as we have mentioned above they may be
related to the dense expanding ring formed around the binary system. The bright knots show 
velocities ranging between the inner and outer shell expansion velocities. This may indicate 
that they are moving with a similar velocity as the particular ambient gas 
in which they are embedded. We may compare these bright knots with the microstructures 
observed in NGC\,2392 \citep[eskimo nebula,][]{ODell02} and NGC\,7662 \citep{PPB04}. Although
each one of these nebulae is seen with a different view angle (NGC\,2392 is seen pole-on and 
NGC\,7662 is seen edge-on) both of them show various low-ionization knots located beyond the 
inner structure and mainly on the outer envelope. If we compare the general morphology of 
NGC\,7354 with that observed in NGC\,2392 and NGC\,7662 one can see that the three of them 
show an ellipsoidal inner structure surrounded by on outer envelope with low-ionization 
bright knots located on the outer structure plus jet-like features, interpreted as FLIERs in 
the case of NGC\,7662. In the case of NGC\,7354 since it is observed with an edge-on view 
angle and it is inclined with respect to the plane of the sky, we are not able to observe the
real morphology of the equatorial brights knots. 
However, one can imagine that if one could get a pole-on view, i.e. with the inner
shell major axis aligned with our line of sight, one would see bright knots located in a zone 
between the inner and outer shell very alike the arragement of knots with tails observed in 
the eskimo nebula. Therefore, having all 
these elements together, we suggest that, as in the cases of NGC\,7662 and NGC\,2392, the 
equatorial bright knots observed in NGC\,7354 are not spherical structures but knots with 
tails, or even filamentary structures. This last suggestion is strongly supported by the high 
resolution [\ion{N}{2}] HST image, Fig.~\ref{fig2}. 

A full model with MHD simulations is beyond the scope of the present study, but we are aware 
that it could clarify some aspects of our work.

\section{Conclusions}

We have carried out a detailed morphological and kinematical analysis of the planetary nebula 
NGC\,7354. In addition, we have derived the physical conditions and elemental abundances in 21
regions of the nebula. The main conclusionas of this work can be sumarized as follows.
 
Physical parameters of all the different structures show that there are slight variations, both
in electron density and electron temperature, within the nebula. Considering our error estimates, 
while density seems to slightly increase inward from the outer shell border to the interior of 
the inner elliptical shell, temperature values seem to slightly increase in the same direction.
Equatorial bright knots are clearly denser that both shells with temperatures in the typical
range for ionized gas. Both jet-like features present a quite low density value and
temperature was only determined for the South jet-like feature. All our derived physical 
parameter values are consistent with the ones expected for radiatively excited gas. Local 
extinction values within the nebula show a slight increasing gradient going from South to 
North of the nebula and from West to East along the location of the equatorial knots.

Based on our kinematical data we have obtained a model that consistently reproduce the 
overall as well as the detailed morphology and PV maps of the nebula, using the 
3D interactive tool {\sc shape}. The final geometrical components included in the model are: 
i) two shells with similar inclination angle (respect to our 
line of sight) but different position angle of the projected semi-major axis;
ii) two semispherical caps placed on the top of the inner elliptical shell;
iii) thirteen spherical knots with different sizes, whose velocities spanned between the 
corresponding velocity values for the outer and inner shells, placed at different positions 
on the outer shell surface according to the optical image and observed spectra; and 
iv) two geometrically thin cylinders corresponding to North and South jet-like features.
 
Finally, although there is no evidence of binarity in NGC\,7354 at present, we suggest a 
qualitative scenario for the formation of the different structures in the nebula
based on the theory of CE evolution and formation of accretion disks in binary 
systems found in the literature.

\acknowledgments

This project has been supported by grants from CONACYT (49002, 45848) and PAPIIT-UNAM 
(IN111903, IN109509). LFM acknowledge support from grants AYA2005-01495 of the Spanish MEC 
(co-funded by FEDER funds), AYA2008-01934 of the Spanish MICINN (co-funded by FEDER funds), 
and FQM1747 of the Junta de Andaluc\'{\i}a. LO aknowledge CONACYT for his posdoctoral research 
scholarship. SZ acknowledge support from the UNAM-ITE colaboration agreement 1500-479-3-V-04.
We are grateful to the staff of all the astronomical facilities and systems used in this 
research, in particular to Mr. Gustavo Melgoza for assistance during OAN-SPM observations.

\clearpage

\begin{figure*}
\epsscale{1.6}
\plotone{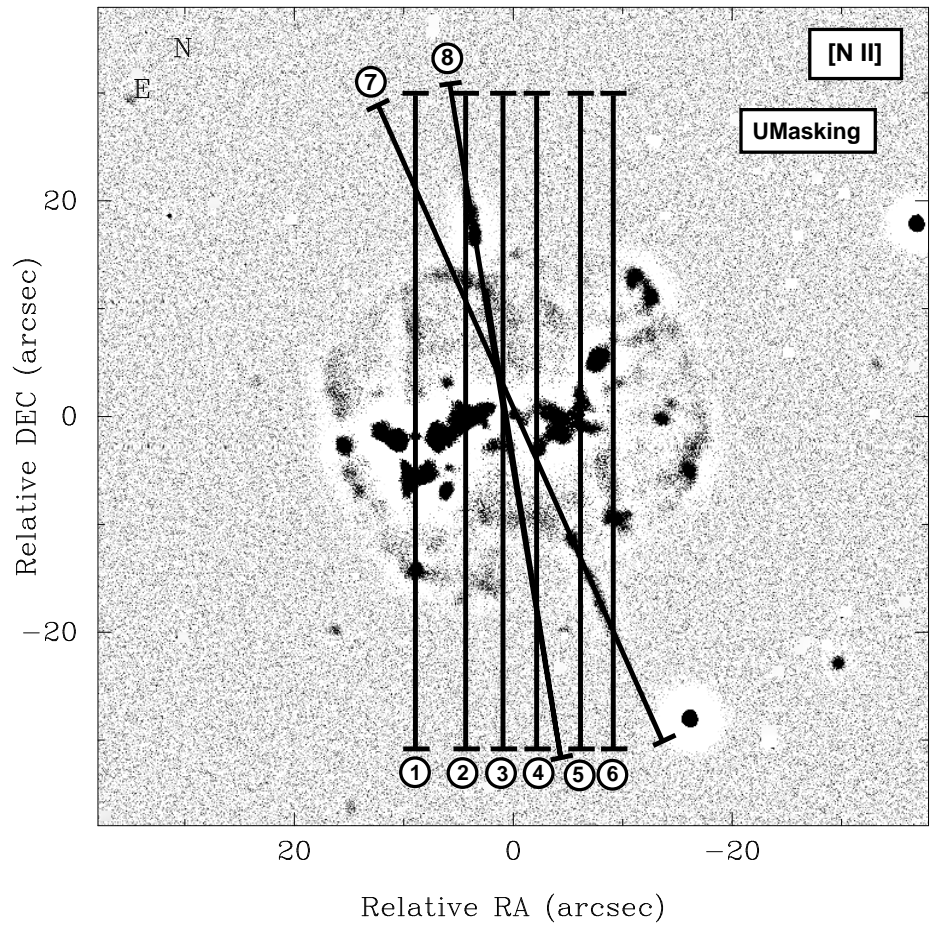}
\caption{Slit positions used for high dispersion spectroscopy superimposed on
the unsharped masking [N\,{\sc ii}] image. The slits are numbered from 1 to 8. 
Slit numbers 7 and 8 are oriented at PAs 25\arcdeg and 13\arcdeg, 
respectively.\label{fig5}}
\end{figure*}

\begin{figure*}
\epsscale{1.8}
\plotone{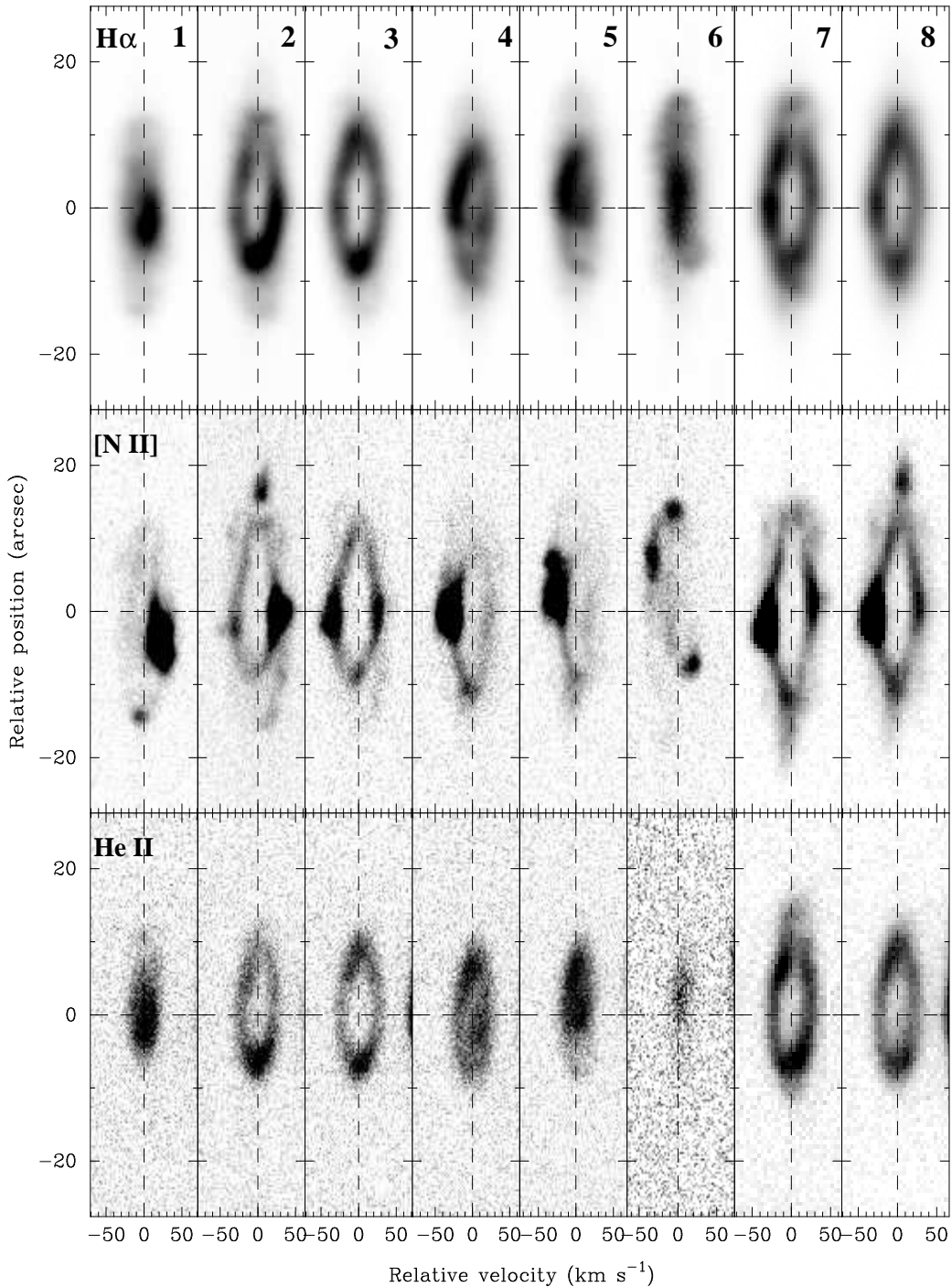}
\caption{Grey-scale position-velocity maps of the H$\alpha$ ({\sl upper panels}), 
[\ion{N}{2}]$\lambda6584$ ({\sl middle panels}) and \ion{He}{2}\,$\lambda6560$ ({\sl lower 
panels}) emission lines derived from the long-slit spectra at slit positions 1 to 8
({\sl indicated in the  H$\alpha$ panels for each column} (see Fig.\,5). The vertical and 
horizontal dashed lines in each panels indicate the systemic velocity of the nebula and the 
declination of the central star, respectively. North is up in each panel.\label{fig6}}
\end{figure*}

\begin{figure*}
\epsscale{1.6}
\plotone{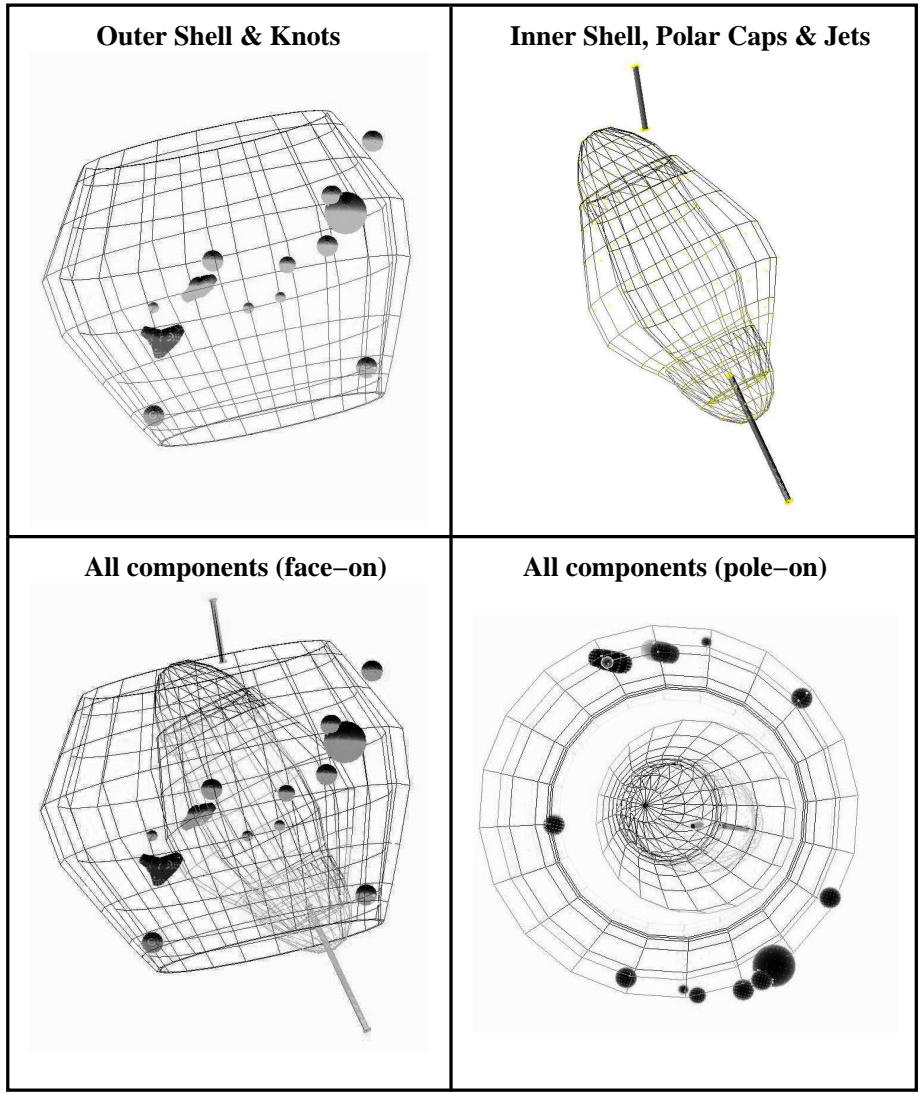}
\caption{Upper panels show two sets of geometrical components used in the 
{\sc shape} model. Lower panels show all components seen from two different 
angles, face-on (lower left panel) and pole-on (lower right panel).\label{fig7}}
\end{figure*}

\begin{figure*}
\epsscale{1.6}
\plotone{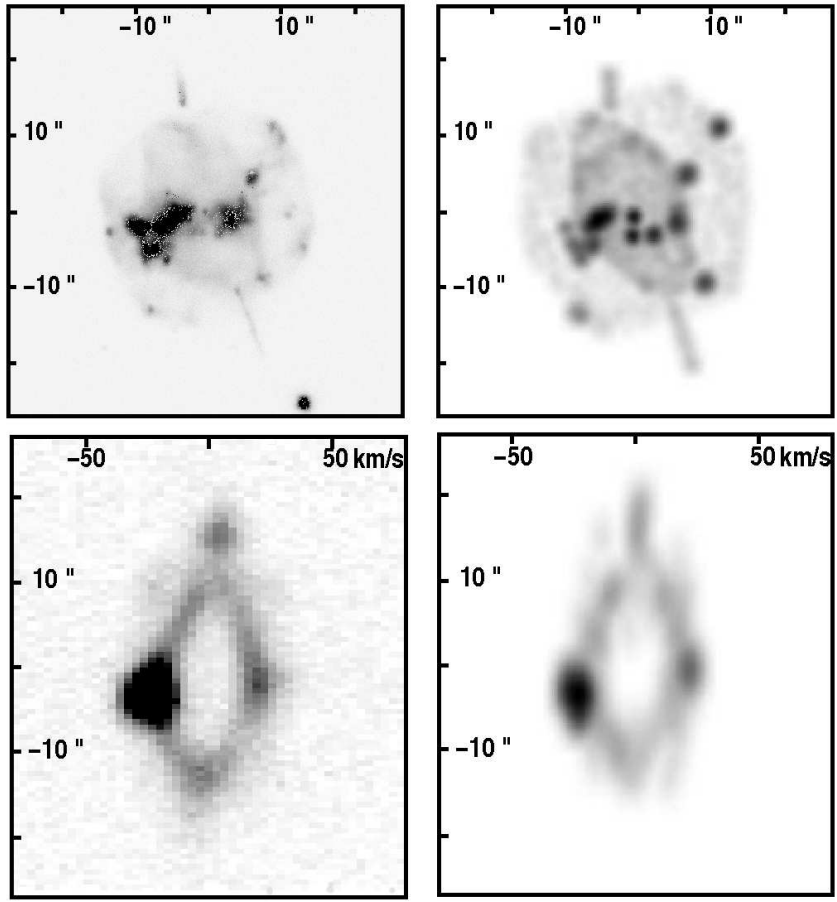}
\caption{Comparison of observed data and the 3D model obtained from {\sc shape}. Upper
panels show the observed [\ion{N}{2}] image (left panel) and the {\sc shape} image (right
panel). Lower panels show the observed PV map of the [\ion{N}{2}]$\lambda6584$ 
line at slit position 7 (left panel) and the corresponding {\sc shape} synthetic PV map 
(right panel).\label{fig8}}
\end{figure*}

\clearpage

\begin{deluxetable}{lrrrrrrrrrrrrr}
\tabletypesize{\scriptsize}
\tablecaption{Dereddened emission-line fluxes (normalized to H$\beta\,=\,100$).\label{tbl-1}}
\tablewidth{0pt}
\tablehead{
\colhead{Ion} & \colhead{$\lambda$} & \colhead{f$_{\lambda}$} & \colhead{A1} & \colhead{A2} &
\colhead{A3} & \colhead{A4} & \colhead{A5} & \colhead{B1} & \colhead{B2} & \colhead{B3} &
\colhead{B4} & \colhead{B5} & \colhead{B6}  
}
\startdata 
H$\gamma$     &    4340 &    0.157 & \nodata &    38.1 &    34.6 &    35.7 &    37.3 &    59.6 &    61.7 &    63.2 &    59.6 &    56.4 &    61.6  \\
{[O III]}     &    4363 &    0.149 & \nodata &    13.5 &    14.1 &    18.3 &    15.4 &    18.7 &    25.1 &    27.3 &    18.8 &    17.3 &    18.2  \\
He I          &    4471 &    0.115 & \nodata &     6.3 &     6.6 & \nodata & \nodata & \nodata & \nodata & \nodata &     6.9 & \nodata &     9.3  \\
He II         &    4686 &    0.050 &     3.8 &    56.3 &    51.6 &    76.9 &    24.8 &     8.4 &    46.9 &    56.9 &    68.1 &    74.9 &    55.5  \\
{[Ar IV]}     &    4711 &    0.042 & \nodata &     8.7 &     7.5 &    11.5 &     7.7 & \nodata &     8.3 &     9.4 &     9.4 &    10.3 &     7.6  \\
{[Ar IV]}     &    4740 &    0.034 & \nodata &     9.3 &     7.1 &    10.9 &     6.6 & \nodata &     6.3 &     5.9 &     7.3 &     7.2 &     6.8  \\
H$\beta$      &    4861 &    0.000 &   100.0 &   100.0 &   100.0 &   100.0 &   100.0 &   100.0 &   100.0 &   100.0 &   100.0 &   100.0 &   100.0  \\
{[O III]}     &    4959 & $-$0.026 &   476.1 &   411.5 &   408.1 &   404.8 &   457.8 &   454.3 &   493.0 &   450.3 &   395.1 &   381.2 &   452.4  \\
{[O III]}     &    5007 & $-$0.038 &  1394.9 &  1216.2 &  1202.8 &  1198.9 &  1358.4 &  1353.1 &  1452.8 &  1323.4 &  1169.0 &  1125.3 &  1333.3  \\
{[N II]}      &    5755 & $-$0.185 & \nodata &     0.5 &     0.9 &     0.3 &     0.8 & \nodata & \nodata &     0.6 & \nodata & \nodata &     1.1  \\
He I          &    5876 & $-$0.203 &    15.4 &     9.8 &    10.9 &     8.2 &    13.7 &    16.3 &    11.4 &     9.7 &     9.0 &     8.5 &    10.7  \\
{[O I]}       &    6300 & $-$0.263 &     5.2 & \nodata &     0.6 &     0.2 &     1.0 & \nodata &     0.4 &     0.4 &     0.2 & \nodata &     0.9  \\
{[S III]}     &    6312 & $-$0.264 & \nodata &     1.8 &     2.3 &     1.6 &     1.5 &     2.8 &     2.3 &     2.2 &     1.7 &     2.1 &     2.1  \\
{[O I]}       &    6364 & $-$0.271 & \nodata & \nodata &     0.2 & \nodata & \nodata & \nodata & \nodata & \nodata & \nodata & \nodata &     0.4  \\
{[N II]}      &    6548 & $-$0.296 &    27.2 &     5.0 &    14.7 &     5.7 &     8.5 &     6.9 &     4.5 &     4.5 &     4.4 &     3.9 &    13.3  \\
H$\alpha$     &    6563 & $-$0.298 &   285.0 &   278.4 &   283.4 &   281.8 &   279.3 &   281.5 &   280.2 &   276.5 &   280.7 &   280.9 &   281.8  \\
{[N II]}      &    6584 & $-$0.300 &    65.7 &    13.9 &    41.4 &    11.7 &    23.8 &    21.8 &    13.2 &    13.0 &    12.2 &    11.3 &    42.3  \\
He I          &    6678 & $-$0.313 &     5.3 &     2.9 &     3.0 &     2.6 &     4.4 &     4.0 &     3.3 &     3.0 &     2.8 &     2.6 &     3.2  \\
{[S II]}      &    6717 & $-$0.318 &     4.6 &     1.3 &     2.7 &     1.0 &     1.8 &     2.7 &     1.2 &     1.2 &     1.1 &     1.1 &     2.3  \\
{[S II]}      &    6731 & $-$0.320 &     5.5 &     1.7 &     3.9 &     1.4 &     2.1 &     2.8 &     1.6 &     1.6 &     1.6 &     1.5 &     3.4  \\
He I          &    7065 & $-$0.364 &     4.0 &     2.7 &     3.2 &     2.3 &     3.5 &     4.1 &     3.2 &     2.7 &     2.6 &     2.2 &     2.6  \\
{[Ar III]}    &    7135 & $-$0.374 &    14.1 &    15.1 &    17.0 &    15.3 &    16.4 &    17.2 &    16.7 &    15.5 &    15.0 &    14.0 &    15.6  \\
{[Ar IV]}     &    7236 & $-$0.387 & \nodata &     1.2 &     1.2 &     1.0 &     1.0 & \nodata &     0.9 &     1.2 &     1.5 &     1.2 &     1.1  \\
He I          &    7281 & $-$0.393 & \nodata &     0.4 &     0.5 &     0.4 &     0.8 & \nodata &     0.6 &     0.6 &     0.6 &     0.4 &     0.6  \\
{[O II]}      &    7320 & $-$0.398 &     1.6 &     1.0 &     1.5 &     1.0 &     1.4 &     1.6 &     0.9 &     1.2 &     1.2 &     1.0 &     1.4  \\
{[O II]}      &    7330 & $-$0.400 &     0.6 &     0.8 &     1.2 &     0.9 &     0.8 & \nodata & \nodata & \nodata & \nodata & \nodata & \nodata \\
&&&&&&&&&&& \\
$c$(H$\beta$)   &       &          &    2.44 &    2.04 &    1.96 &    1.86 &    1.73 &    2.05 &    2.12 &   2.12  &    1.98 &    1.99 &   2.03  \\
log $I$(H$\beta$) &     &          & -16.873 & -14.998 & -14.898 & -15.040 & -15.016 & -15.901 & -15.891 & -15.461 & -15.235 & -15.311 & -15.417  \\
\enddata
\end{deluxetable}

\begin{deluxetable}{lrrrrrrrrrrrr}
\tabletypesize{\scriptsize}
\tablecaption{---Continued}
\tablewidth{0pt}
\tablenum{1}
\tablehead{
\colhead{Ion} & \colhead{$\lambda$} & \colhead{f$_{\lambda}$} & 
\colhead{B7} & \colhead{C1} &\colhead{C2} & \colhead{C3} & \colhead{C4} & \colhead{C5} & \colhead{C6} & 
\colhead{D1} &\colhead{D2} & \colhead{D3} 
}
\startdata
H$\gamma$     &    4340 &    0.127 &    60.3 &   \nodata &    58.4 &    69.6  &    48.9 &    49.2 &    55.8 &     77.9 &    57.7 &    60.9   \\
{[O III]}     &    4363 &    0.121 &    23.7 &   \nodata &     9.7 &  \nodata &    14.5 &    23.1 &    13.0 &     15.4 & \nodata &    15.0   \\
He I          &    4471 &    0.095 & \nodata &   \nodata &     7.6 &  \nodata & \nodata & \nodata & \nodata &  \nodata & \nodata & \nodata   \\
He II         &    4686 &    0.043 &     8.6 &       1.4 &    34.8 &    39.9  &    58.0 &    63.1 &    22.7 &      6.7 &     7.3 &     9.0   \\
{[Ar IV]}     &    4711 &    0.037 &     6.2 &   \nodata &    11.1 &    13.7  &     8.3 &     8.6 & \nodata &  \nodata & \nodata &   \nodata \\
{[Ar IV]}     &    4740 &    0.030 & \nodata &   \nodata &     2.7 &  \nodata &     6.5 &     8.3 &     7.4 &  \nodata & \nodata &   \nodata \\
H$\beta$      &    4861 & $-$0.000 &   100.0 &     100.0 &   100.0 &   100.0  &   100.0 &   100.0 &   100.0 &    100.0 &   100.0 &   100.0   \\
{[O III]}     &    4959 & $-$0.024 &   461.4 &     402.9 &   464.1 &   457.7  &   403.1 &   381.6 &   475.1 &    465.4 &   449.5 &   478.3   \\
{[O III]}     &    5007 & $-$0.036 &  1380.8 &    1193.7 &  1364.1 &  1352.5  &  1181.0 &  1127.9 &  1416.6 &   1384.9 &  1334.6 &  1403.1   \\
{[N II]}      &    5755 & $-$0.195 & \nodata &       1.2 &     1.6 &     1.3  &     1.5 &     0.8 & \nodata &  \nodata & \nodata &  \nodata  \\
He I          &    5876 & $-$0.215 &    16.2 &      16.7 &    12.2 &    10.5  &    10.3 &     8.8 &    14.1 &     16.2 &    16.0 &    15.8   \\
{[O I]}       &    6300 & $-$0.282 &     0.4 &       1.6 &     2.6 &     1.8  &     2.4 &     0.8 &     0.3 &      0.4 & \nodata &     1.0   \\
{[S III]}     &    6312 & $-$0.283 &     2.1 &       1.8 &     2.2 &     2.3  &     2.3 &     2.0 &     1.9 &      1.5 &     2.2 &     1.5   \\
{[O I]}       &    6364 & $-$0.291 & \nodata &   \nodata &     0.9 &     1.0  &     0.7 &     0.2 & \nodata &  \nodata & \nodata &  \nodata  \\
{[N II]}      &    6548 & $-$0.318 &     8.0 &      20.8 &    28.2 &    22.1  &    26.8 &    10.5 &     7.0 &     18.4 &     5.8 &     8.8   \\
H$\alpha$     &    6563 & $-$0.320 &   280.2 &     284.2 &   283.5 &   283.5  &   283.9 &   282.0 &   283.5 &    282.7 &   285.0 &   283.0   \\
{[N II]}      &    6584 & $-$0.323 &    24.5 &      65.2 &    82.8 &    66.6  &    82.8 &    31.8 &    21.0 &     58.4 &    19.5 &    28.2   \\
He I          &    6678 & $-$0.336 &     4.1 &       4.4 &     3.5 &     3.5  &     3.2 &     2.7 &     4.2 &      4.6 &     4.5 &     5.1   \\
{[S II]}      &    6717 & $-$0.342 &     1.9 &       3.5 &     3.6 &     2.8  &     3.4 &     1.8 &     2.2 &      4.2 &     1.9 &     2.1   \\
{[S II]}      &    6731 & $-$0.344 &     2.5 &       5.6 &     5.0 &     4.1  &     5.1 &     2.7 &     2.2 &      5.3 &     2.5 &     3.4   \\
He I          &    7065 & $-$0.387 &     3.6 &       4.2 &     3.5 &     3.3  &     3.3 &     2.6 &     3.3 &      4.2 &     4.5 &     3.7   \\
{[Ar III]}    &    7135 & $-$0.396 &    16.0 &      19.6 &    18.0 &    17.3  &    17.2 &    15.2 &    14.5 &     18.7 &    15.9 &    15.8   \\
{[Ar IV]}     &    7236 & $-$0.409 &     0.5 &   \nodata &     0.9 &     1.1  &     1.0 &     1.3 & \nodata &  \nodata & \nodata &  \nodata  \\ 
He I          &    7281 & $-$0.414 &     0.7 &   \nodata &     0.6 &     0.5  &     0.4 &     0.6 & \nodata &  \nodata &     1.4 &     0.4   \\
{[O II]}      &    7320 & $-$0.419 &     1.0 &       1.6 &     2.3 &     2.0  &     1.9 &     1.2 &     0.5 &      2.9 &     1.1 &     0.6   \\
{[O II]}      &    7330 & $-$0.420 & \nodata &       1.4 &     2.1 &     2.1  &     2.3 &     1.3 &     1.4 &  \nodata & \nodata & \nodata   \\
&&&&&&&&&&& \\
$c$(H$\beta$)   &         &        &    2.01 &      2.20 &    2.41 &     2.24 &    1.95 &    2.02 &    1.71 &    1.78 &    1.83 &    1.83   \\ 
log $I$(H$\beta$) &     &          & -15.801 &   -16.357 & -16.056 &  -16.224 & -15.203 & -15.313 & -15.399 & -15.967 & -15.874 & -15.913 
\enddata
\end{deluxetable}

\begin{deluxetable}{lccccccc}
\tabletypesize{\footnotesize}
\tablewidth{0pt}
\tablecaption{Derived physical parameters.\label{tbl-2}}
\tablehead{
\colhead{Region}           &  &
\multicolumn{2}{c}{$T_{\rm_e}$ [K]} & &
\multicolumn{3}{c}{$N_{\rm_e}$ [cm$^{-3}$]} \\[0.5ex]
\tableline \\[-1.5ex]
&&\colhead{[\ion{N}{2}]}      &   \colhead{[\ion{O}{3}]}  & & 
\colhead{[\ion{S}{2}]}        &  \colhead{[\ion{Cl}{3}]}   &
\colhead{[\ion{Ar}{4}]} 
}
\startdata
A1   &&  \nodata          & \nodata        &  & 1141$\pm$1080 & \nodata         & \nodata    \\
A2   &&  15909$\pm$1149   & 12040$\pm$243  &  & 2048$\pm$350  & 5545$\pm$ 2660  & 5658$\pm$1319  \\
A3   &&  11459$\pm$386    & 12281$\pm$262  &  & 2623$\pm$281  & 4023$\pm$ 1888  & 3754$\pm$1436  \\
A4   &&  12870$\pm$1404   & 13618$\pm$322  &  & 1982$\pm$557  & 1033$\pm$952    & 3812$\pm$1201  \\ 
A5   &&  15125$\pm$1920   & 12135$\pm$447  &  & 1404$\pm$458  & \nodata         & 2341$\pm$1587     \\
\tableline
B1   &&  \nodata          & 13130$\pm$700  &  & 751$\pm$267   & \nodata         & \nodata    \\
B2   &&  \nodata          & 14346$\pm$519  &  & 1799$\pm$610  & \nodata         & \nodata    \\
B3   &&  17690$\pm$1468   & 15523$\pm$356  &  & 2402$\pm$515  & \nodata         & \nodata    \\
B4   &&  \nodata          & 13885$\pm$263  &  & 3106$\pm$555  & 7708$\pm$4389   & 1062$\pm$767 \\
B5   &&  \nodata          & 13646$\pm$286  &  & 2216$\pm$415  & 15086$\pm$9695  & \nodata    \\
B6   &&  12833$\pm$567    & 13005$\pm$298  &  & 2700$\pm$356  & \nodata         & 2986$\pm$1423  \\
B7   &&  \nodata          & 14334$\pm$956  &  & 1910$\pm$636  & \nodata         & \nodata    \\
\tableline
C1   &&  10724$\pm$2902   & \nodata        &  & 3451$\pm$4044 & \nodata         & \nodata    \\
C2   &&  11312$\pm$785    & 10363$\pm$1162 &  & 2130$\pm$629  & \nodata         & \nodata    \\
C3   &&  11329$\pm$1438   & \nodata        &  & 2354$\pm$1194 & \nodata         & \nodata    \\
C4   &&  11006$\pm$481    & 12478$\pm$755  &  & 2715$\pm$551  & \nodata         & \nodata    \\
C5   &&  12662$\pm$1271   & 15454$\pm$1039 &  & 3311$\pm$1290 & \nodata         & \nodata    \\
C6   &&   \nodata         & 11298$\pm$1564 &  &  664$\pm$910  & \nodata         & \nodata    \\
\tableline
D1   &&  \nodata          & 12066$\pm$984  &  & 1603$\pm$575  & \nodata         & \nodata    \\
D2   &&  \nodata          & \nodata        &  & 1574$\pm$725  & \nodata         & \nodata    \\
D3   &&  \nodata          & 11825$\pm$835  &  & 4102$\pm$2481 & \nodata         & \nodata   
\enddata
\end{deluxetable}

\begin{deluxetable}{lccccccccccc}
\tabletypesize{\scriptsize}
\tablewidth{0pt}
\tablecaption{Derived ionic abundances.\label{tbl-3}}
\tablehead{
\colhead{Region}           & \multicolumn{11}{c}{Ion} \\[0.5ex]
\tableline \\[-1.5ex]
& \colhead{He$^+$}  & \colhead{He$^{++}$} & \colhead{N$^+$} & \colhead{O$^0$} & 
  \colhead{O$^{+1}$}  & \colhead{O$^{+2}$}  & \colhead{S$^+$} & \colhead{S$^{+2}$} & 
  \colhead{Ar$^{+2}$} & \colhead{Ar$^{+3}$} & \colhead{Ar$^{+4}$} \\
&  &  & [1$\times 10^{-5}$] & [1$\times 10^{-6}$] & [1$\times 10^{-5}$] & [1$\times 10^{-4}$] 
      & [1$\times 10^{-7}$] & [1$\times 10^{-6}$] & [1$\times 10^{-6}$] & [1$\times 10^{-6}$] 
      & [1$\times 10^{-7}$] 
}
\startdata
A1   &  0.117 &  0.003  &   1.38  &  10.12  &   2.98   &   4.89 &   2.83 & \nodata &   1.30  & \nodata & \nodata \\
A2   &  0.073 &  0.046  &   0.10  & \nodata &   0.18   &   1.10 &   0.37 &   0.80  &   0.55  &   0.49  &   0.69 \\ 
A3   &  0.074 &  0.042  &   0.60  &   0.74  &   1.11   &   2.69 &   1.63 &   3.17  &   1.15  &   0.91  &   1.00 \\
A4   &  0.058 &  0.062  &   0.13  &   0.15  &   0.49   &   1.90 &   0.42 &   1.47  &   0.82  &   1.04  &   1.06 \\ 
A5   &  0.099 &  0.020  &   0.19  &   0.50  &   0.28   &   1.40 &   0.47 &   0.81  &   0.65  &   0.46  & \nodata \\ 
B1   &  0.112 &  0.007  &   0.42  & \nodata &    3.98  &   4.76 &   1.45 &   6.68  &   1.59  & \nodata & \nodata \\
B2   &  0.082 &  0.038  &   0.26  &   0.72  &    1.60  &   5.15 &   0.87 &   5.54  &   1.54  &   1.32  & \nodata \\ 
B3   &  0.062 &  0.048  &   0.08  &   0.11  &    0.13  &   0.94 &   0.30 &   0.74  &   0.48  &   0.37  &   0.39 \\ 
B4   &  0.069 &  0.055  &   0.25  &   0.35  &    1.86  &   4.14 &   0.99 &   3.99  &   1.39  &   1.54  &   2.50 \\ 
B5   &  0.061 &  0.061  &   0.23  & \nodata &    1.71  &   3.98 &   0.82 &   4.88  &   1.29  &   1.60  &   2.90 \\ 
B6   &  0.074 &  0.045  &   0.46  &   0.86  &    0.57  &   2.13 &   1.12 &   1.89  &   0.84  &   0.67  &   0.85 \\ 
B7   &  0.110 &  0.007  &   0.48  &   0.74  &    1.88  &   4.84 &   1.39 &   4.89  &   1.48  &   1.03  & \nodata \\ 
C1   &  0.110 &  0.001  &   1.11  &   2.46  &   1.56   &   3.29 &   2.87 &   3.08  &   1.53  & \nodata & \nodata \\ 
C2   &  0.083 &  0.028  &   1.21  &   3.22  &   2.13   &   3.08 &   2.06 &   3.22  &   1.26  &   1.32  &   0.88 \\ 
C3   &  0.088 &  0.032  &   0.98  &   3.41  &   1.96   &   3.14 &   1.69 &   3.27  &   1.20  &   1.64  &   1.11 \\ 
C4   &  0.080 &  0.047  &   1.31  &   3.18  &   2.46   &   3.01 &   2.35 &   3.69  &   1.27  &   1.02  &   1.07 \\ 
C5   &  0.068 &  0.051  &   0.36  &   0.63  &   0.60   &   1.89 &   0.98 &   1.92  &   0.84  &   0.81  &   1.64 \\ 
C6   &  0.101 &  0.018  &   0.41  &   0.67  &   1.34   &   4.98 &   1.15 &   4.69  &   1.33  &   1.56  & \nodata \\ 
D1   &  0.116 &  0.005  &   1.13  &   0.69  &   5.51   &   4.88 &   2.85 &   3.51  &   1.73  & \nodata & \nodata \\
D2   &  0.113 &  0.006  &   0.38  & \nodata &   2.18   &   4.65 &   1.31 &   5.21  &   1.46  & \nodata & \nodata \\ 
D3   &  0.113 &  0.007  &   0.57  &   1.88  &   0.87   &   4.92 &   2.16 &   3.51  &   1.45  & \nodata & \nodata 
\enddata
\tablecomments{In regions A5, B1 and D3, He$^{++}$ abundances shown are an upper limit.}
\end{deluxetable}

\begin{deluxetable}{lccccc}
\tabletypesize{\scriptsize}
\tablewidth{0pt}
\tablecaption{Derived total abundances.\label{tbl-4}}
\tablehead{
\colhead{Region}  & \colhead{He/H} &  \colhead{O/H}  &   \colhead{N/H}  &   \colhead{S/H} & \colhead{Ar/H}
}
\startdata 
   A1 &   0.120$\pm$0.010 &   8.72$\pm$0.05 & 8.39$\pm$0.35 & 6.67$\pm$1.54 & 6.38$\pm$0.10 \\ 
   A2 &   0.118$\pm$0.001 &   8.19$\pm$0.14 & 7.94$\pm$0.28 & 6.41$\pm$0.14 & 6.05$\pm$0.01 \\
   A3 &   0.116$\pm$0.001 &   8.58$\pm$0.15 & 8.31$\pm$0.21 & 6.88$\pm$0.09 & 6.35$\pm$0.01 \\ 
   A4 &   0.121$\pm$0.001 &   8.50$\pm$0.40 & 7.94$\pm$0.60 & 6.63$\pm$0.27 & 6.30$\pm$0.01 \\ 
   A5 &   0.119$\pm$0.001 &   8.21$\pm$0.28 & 8.02$\pm$0.59 & 6.36$\pm$0.28 & 6.09$\pm$0.14 \\ 
   B1 &   0.118$\pm$0.002 &  8.73$\pm$0.11 &  7.76$\pm$0.13 & 7.06$\pm$0.05 & 6.47$\pm$0.10 \\
   B2 &   0.120$\pm$0.002 &  8.84$\pm$0.08 &  8.05$\pm$0.12 & 7.14$\pm$0.04 & 6.46$\pm$0.10 \\
   B3 &   0.110$\pm$0.001 &  8.15$\pm$0.18 &  7.90$\pm$0.26 & 6.40$\pm$0.14 & 5.95$\pm$0.01 \\
   B4 &   0.124$\pm$0.001 &  8.81$\pm$0.04 &  7.94$\pm$0.07 & 6.97$\pm$0.03 & 6.51$\pm$0.01 \\
   B5 &   0.122$\pm$0.001 &  8.82$\pm$0.05 &  7.94$\pm$0.07 & 7.07$\pm$0.03 & 6.51$\pm$0.01 \\
   B6 &   0.119$\pm$0.001 &  8.48$\pm$0.15 &  8.38$\pm$0.21 & 6.72$\pm$0.10 & 6.21$\pm$0.01 \\
   B7 &   0.117$\pm$0.002 &  8.72$\pm$0.16 &  8.13$\pm$0.18 & 7.03$\pm$0.07 & 6.44$\pm$0.10 \\
   C1 &   0.111$\pm$0.006 &  8.54$\pm$0.89 &  8.39$\pm$4.87 & 6.82$\pm$1.90 & 6.46$\pm$0.40 \\ 
   C2 &   0.111$\pm$0.003 &  8.60$\pm$0.30 &  8.36$\pm$0.45 & 6.81$\pm$0.20 & 6.45$\pm$0.03 \\ 
   C3 &   0.120$\pm$0.006 &  8.61$\pm$0.27 &  8.31$\pm$0.74 & 6.83$\pm$0.36 & 6.49$\pm$0.04 \\ 
   C4 &   0.127$\pm$0.002 &  8.65$\pm$0.26 &  8.37$\pm$0.33 & 6.86$\pm$0.14 & 6.41$\pm$0.02 \\ 
   C5 &   0.120$\pm$0.003 &  8.45$\pm$0.55 &  8.23$\pm$0.72 & 6.71$\pm$0.29 & 6.27$\pm$0.02 \\ 
   C6 &   0.120$\pm$0.005 &  8.76$\pm$0.23 &  8.25$\pm$0.97 & 7.07$\pm$0.33 & 6.40$\pm$0.10 \\ 
   D1 &   0.122$\pm$0.007 &  8.75$\pm$0.14 &  8.06$\pm$0.16 & 6.77$\pm$0.08 & 6.51$\pm$0.10 \\  
   D2 &   0.118$\pm$0.002 &  8.70$\pm$0.01 &  7.94$\pm$0.13 & 7.03$\pm$0.06 & 6.44$\pm$0.10 \\ 
   D3 &   0.120$\pm$0.002 &  8.72$\pm$0.13 &  8.53$\pm$0.17 & 7.01$\pm$0.09 & 6.43$\pm$0.10 \\[0.5ex]
\tableline \\[-1.5ex]
PNe$^{a}$    & 0.12$\pm$0.02 & 8.68$\pm$0.15 & 8.35$\pm$0.25 & 6.92$\pm$0.30 & 6.39$\pm$0.30  \\
TI PNe$^{b}$ & 0.13$\pm$0.04 & 8.65$\pm$0.15 & 8.72$\pm$0.15 & 6.91$\pm$0.30 & 6.42$\pm$0.30  \\
H II$^{c}$   & 0.10$\pm$0.01 & 8.70$\pm$0.04 & 7.57$\pm$0.04 & 7.06$\pm$0.06 & 6.42$\pm$0.04  \\
Sun$^{d}$    & 0.09$\pm$0.01 & 8.66$\pm$0.05 & 7.78$\pm$0.06 & 7.14$\pm$0.05 & 6.18$\pm$0.08  
\enddata
\tablecomments{Except for He, all abundances relative to H are logarithmic values with logH\,=\,+12.
Final rows are for comparison: (a) Average for PNe \citep{KB94}; (b) Average for 
Type I PNe \citep{KB94}; (c) Average for H II regions \citep{Sh83}; the Sun 
\citep{GAS07}. In regions A5, B1, and D3, He abundances shown are an upper limit.}
\end{deluxetable}

\begin{deluxetable}{lccccc}
\tabletypesize{\scriptsize}
\tablewidth{0pt}
\tablecaption{Geometrical and Kinematic Parameters from our {\sl SHAPE} model.\label{tbl-5}}
\tablehead{
\colhead{Geometrical Structure}  & \colhead{Size} &  \colhead{Expansion Velocity}  &  
\colhead{Inclination Angle $i$} &  \colhead{Position Angle $\phi$} & \colhead{Kinematical Age}\\
& \colhead{[$''$]}  &  \colhead{[km s$^{-1}$]} &  \colhead{[deg]} & \colhead{[deg]} & \colhead{[yr]}
}
\startdata 
Outer Elliptical Shell &  16 $\times$ 15 &  45  &  94 &  18 & 2 500 \\ 
Inner Elliptical Shell &  10 $\times$ 8  &  35  &  94 &  28 & 1 600 \\
Semispherical cap  N  &  14  &  47  &  94 &  28 &  1 000\\
Semispherical cap  S  &  14  &  47  &  85 &  203 &  1 000\\
Spherical Knots &  1 to 2 & 25 to 44  &  \nodata & \nodata\\
Thin Cylinder  N  &  7  &  60  & 95  & 3  &  1 900 \\
Thin Cylinder  S  &  13  &  60  &  85  & 205 & 1 900
\enddata
\tablecomments{Size refers to major and minor axis in the case of elliptical structures,
length for thin cylinders, and diameter in the case of the spherical knots and polar caps.
All kinematical ages were determined assuming a distance to the nebula of 1.5 kpc \citep{S86}.}
\end{deluxetable}


\begin{thebibliography}{}
\bibitem[Akashi \& Soker(2008)]{AS08} Akashi, M. \& Soker, N. 2008, \mnras, 391, 1063
\bibitem[Balick(1987)]{B87} Balick, B. 1987, \aj, 94, 671
\bibitem[Balick et al.(1987)]{BPI87} Balick, B., Preston, H.L. \& Icke, V. 1987,
   \aj, 94, 1641
\bibitem[Balick \& Frank(2002)]{BF02} Balick, B. \& Frank, A. 2002, \araa, 40, 439
\bibitem[Cahn et al.(1992)]{CKS92} Cahn, J.H., Kaler, J.B. \& Stanghellini,
   L. 1992, \aaps, 94, 399 
\bibitem[Cardelli, Clayton \& Mathis(1989)]{CCM89} Cardelli, J.A., Clayton, G.C.
   \& Mathis, J.S.  1989, \apj, 345, 245
\bibitem[Daub(1982)]{D82} Daub, C.T. 1982, \apj, 260, 612
\bibitem[Dennis et al.(2008)]{D08} Dennis, T.J., Cunningham, A.J., Frank, A., Balick, B.,
   Blackman, E.G. \& Mitran, S. 2008, \apj, 679, 1327
\bibitem[DeRobertis, Dufour \& Hunt (1987)]{dRDH87} DeRobertis, M.M., Dufour, R. 
   \& Hunt, R. 1987, \jrasc, 81, 195 
\bibitem[Frank et al.(1996)]{F96} Frank, A., Balick, B. \& Livio, M. 1996, \apj, 471, L53
\bibitem[Garc\'\i a-Arredondo \& Frank (2004)]{GAF04} Garc\'\i a-Arredondo, F. \& Frank, A. 
   2004, \apj, 600, 992
\bibitem[Grevesse, Asplund \& Sauval (2007)]{GAS07} Grevesse, N., Asplund, M. \& Sauval, A.J. 
   2007, \ssr, 130, 105
\bibitem[Guerrero et al. (2008)]{G08} Guerrero, M.A., Miranda, L.F., Riera, A., Vel\'azquez, 
   P.F., Olgu\'\i n, L., V\'azquez, R., Chu, Y.-H., Raga, A., Ben\'\i tez, G. 2008, \apj, 683, 
   272
\bibitem[Hajian et al. (1997)]{H97} Hajian, A.R., Balick, B., Terzian, Y. \& Perinotto, M. 1997
   \apj, 487, 304
\bibitem[Hajian et al. (2007)]{H07} Hajian, A.R., Movit, S.M., Trofimov, D., Balick, B., 
  Terzian, Y., Knuth, K.H., Granquist-Fraser, D., Huyser, K.A., Jalobeanu, A., McIntosh, D.,
  Jaskot, A.E., Palen, S. \& Panagia, N. 2007, \apjs, 169, 289 
\bibitem[Kingsburgh \& Barlow (1994)]{KB94} Kingsburgh, R.L. \& Barlow, M.J. 1994,\mnras, 271,257
\bibitem[Kwok et al. (1978)]{K78} Kwok, S., Purton, C.R. \& Fitzgerald, P.M.
   1978, \apj, 219, L125
\bibitem[Lee \& Sahai (2003)]{LS03} Lee, C.-F. \& Sahai, R. 2003, \apj, 586, 319
\bibitem[L\'opez et al. (2000)]{L00} L\'opez, J. A., Meaburn, J., Rodríguez, L. F., V\'azquez,
   R., Steffen, W., Bryce M. 2000, \apj, 538, 233
\bibitem[Manchado et al. (1996)]{M96} Manchado, A., Guerrero, M.A., Stanghellini, L. \& Serra-
   Ricart, M. 1996, The IAC morpholigical catalog of northern Galactic planetary nebulae, Pub. La 
   Laguna, Spain:Instituto Astrof\'\i sico de Canarias
\bibitem[Martins \& Viegas (2002)]{MV02} Martins, L.P. \& Viegas, S.M. 2002, \aap, 387, 1074
\bibitem[Meaburn et al.(2003)]{M03} Meaburn, J., L\'opez, J.A., Guti\'errez, L. et al. 2003,
   \rmxaa, 39, 185
\bibitem[Mellema (1995)]{M95} Mellema, G. 1995, \mnras, 277, 173
\bibitem[Mellema \& Frank (1995)]{MF95}  Mellema, G. \& Frank, A. 1995, \mnras, 273, 401
\bibitem[Mezger \& Henderson (1967)]{MH67} Mezger, P.G. \& Henderson, A.P. 1967, \apj, 147, 471
\bibitem[Miranda \& Solf (1992)]{MS92} Miranda, L.F. \& Solf, J. 1992, \aap, 260, 397
\bibitem[Miranda et al. (2006)]{MAVG06} Miranda, L.F., Ayala, S.,  V\'azquez, R. \& Guill\'en, 
   P.F. 2006, \aap, 456, 591
\bibitem[Miranda, Pereira \& Guerrero (2009)]{MPG09} Miranda, L.F., Pereira, C.B. \& Guerrero,
   M.A. 2009, \aj, 137, 414
\bibitem[O'Dell et al. (2002)]{ODell02} O'Dell, C.R., Balick, B., Hajian, A.R., Henney, W.J. 
   \& Burkert, A. 2002, \aj, 123, 3329 
\bibitem[Perinotto et al. (2004a)]{P04} Perinotto, M., Morbidelli, L. \& Scatarzi, 
   A. 2004a, \mnras, 349, 793
\bibitem[Perinotto et al. (2004b)]{PPB04} Perinotto, M., Patriarchi, P., Balick, B. \& Corradi, 
   R.L.M. 2004b, \aap, 422, 963
\bibitem[Phillips (2004)]{Ph04} Phillips, J.P. 2004, \mnras, 353, 589
\bibitem[Phillips et al. (2009)]{Ph09} Phillips, J.P., Ramos-Larios, G., Schr\"oeder, K.-P.
   \& Contreras, J.L. Verbena 2009, \mnras, 399, 1126
\bibitem[Rasio \& Livio (1996)]{RL96} Rasio, F.A. \& Livio, M. 1996, \apj, 4
   71, 366
\bibitem[Sabbadin (1986)]{S86} Sabbadin, F. 1986, \aaps, 64, 579
\bibitem[Sabbadin et al. (1983)]{S83} Sabbadin, F., Bianchini, A. \&
   Hamzaoglu, E. 1983, \aap, 51, 119
\bibitem[Sahai \& Trauger (1998)]{ST98} Sahai, R. \& Trauger, John T. 1998, \aj, 116, 1357
\bibitem[Sahai et al. (2007)]{S07} Sahai, R., Morris, M., S\'anchez Contreras, C. \& Claussen,
   M. 2007, \aj, 134, 2200 
\bibitem[Sandquist et al. (1998)]{S98} Sandquist, E.L., Taam, R.E., Chen, X., Bodenheimer, P. 
   \& Burkert, A. 1998, \apj, 500, 909
\bibitem[Schwarz et al. (1992)]{SCM92} Schwarz, H.E., Corradi, R. \& Melnick, J. 
  1992, \aaps, 96, 23 
\bibitem[Shaver et al. (1983)]{Sh83} Shaver, P.A., McGee, R.X., Newton, L.M., Danks, A.C. 
   \& Pottasch, S.R. 1983, \mnras, 204,53
\bibitem[Shaw \& Dufour (1994)]{SD94} Shaw, R.A. \& Dufour, R.J. 1994, in D.R. Crabtree, Hanisch, 
   R.J., Barnes, J., eds., ASP Conf. Ser. Vol. 61, {\sl Astronomical Data Analysis Software and 
   Systems III.} Astron. Soc. Pac., San Francisco, p.327
\bibitem[Soker \& Livio (1994)]{SL94} Soker, N. \& Livio, M. 1994, \apj, 421, 219
\bibitem[Soker \& Rappaport (2000)]{SR00} Soker, N. \& Rappaport, S. 2000, \apj, 538, 241
\bibitem[Stanghellini et al. (2006)]{S06} Stanghellini, L., Guerrero, M.A., Cunha, K., Manchado, 
  A. \& Villaver, E. 2006, \apj, 651, 898
\bibitem[Steffen \& L\'opez (2006)]{SL06} Steffen, W. \& L\'opez, J.A. 2006, \rmxaa, 42, 99
\bibitem[Terman \& Taam (1996)]{TT96} Terman, J.L. \& Taam, R.E. 1996, \apj, 458, 692
\bibitem[Terzian et al. (1974)]{T74} Terzian, Y., Balick, B., \& Bignell, C. 1974, \apj, 188, 257
\bibitem[V\'azquez et al. (1998)]{V98} V\'azquez R., Kingsburgh R.L., L\'opez J.A. 1998, \mnras, 
  296, 564
\bibitem[V\'azquez et al. (1999)]{V99} V\'azquez R., L\'opez J.A., Miranda, L.F., Torrelles,
  J.M., Meaburn, J. 1999, \mnras, 308, 939
\bibitem[V\'azquez et al. (2008)]{V08} V\'azquez R., Miranda, L.F., Olgu\'\i n, L., Ayala, S., 
  Torrelles, J.M., Contreras, M.E. and Guill\'en, P.F. 2008, \aap, 481, 107
\bibitem[Zhang (1995)]{Z95} Zhang, C.Y. 1995, \apjs, 98, 659
\end{thebibliography}
\end{document}